# Enlarging the class of exactly solvable nonrelativistic problems

A. D. Alhaidari

*Physics Department, King Fahd University of Petroleum & Minerals, Box 5047, Dhahran 31261, Saudi Arabia*
e-mail: haidari@mailaps.org

We lift the constraint of a diagonal representation of the Hamiltonian by searching for square integrable bases that support a tridiagonal matrix representation of the wave operator. Doing so results in exactly solvable problems with a class of potentials which is larger than, and/or generalization of, what is already known. In addition, we found new representations for the solution space of some well known potentials. The problem translates into finding solutions of the resulting three-term recursion relation for the expansion coefficients of the wavefunction. Some of these solutions are new kind of orthogonal polynomials. The examples given, which are not exhaustive, are for problems in one and three dimensions. The analytic solutions obtained by this approach include the discrete as well as the continuous spectrum of the Hamiltonian.

PACS numbers: 03.65.Ge, 02.30.Gp, 03.65.Db, 03.65.Ca

## I. INTRODUCTION

Exact solutions of the wave equation are important because of the conceptual understanding of physics that can only be brought about by such solutions. Moreover, these solutions are valuable means for checking and improving models and numerical methods being introduced for solving complicated physical problems. In fact, in some limiting cases or for some special circumstances they may constitute analytic solutions of realistic problems or approximations thereof. Most of the known exactly solvable problems fall within distinct classes of shape invariant potentials [1]. Supersymmetric quantum mechanics [2], potential algebras [3], and point canonical transformations [4] are three methods among many which are used in the search for exact solutions of the wave equation. In nonrelativistic quantum mechanics, this development was carried out over the years by many authors where several classes of shape invariant potentials being accounted for and tabulated (see, for example, the references cited in [1]). It was also extended to other classes of conditionally exactly [5] and quasi exactly [6] solvable problems where all or, respectively, part of the energy spectrum is known. Recently, the relativistic extension of some of these formulations was carried out where several relativistic problems where formulated and solved exactly. These include, but not limited to: the Dirac-Morse, Dirac-Scarf, Dirac-Pöschl-Teller, Dirac-Hulthén...etc. [7]. A superalgebra which is a special graded extension of so(2,1) Lie algebra was found to be associated with some of these classes of exactly solvable relativistic problems [8].

In all of these developments, the main objective is to find solutions of the eigenvalue wave equation $H|\psi\rangle = E|\psi\rangle$, where $H$ is the self-adjoint Hamiltonian and $E$ is the real energy which is either discrete (for bound states) or continuous (for scattering states). In most cases, especially in the search for algebraic or numerical solutions, the wave function $\psi$ spans the space of square integrable functions with discrete basis elements $\{f_n\}_{n=0}^{\infty}$. That is, the wavefunction is expandable as $|\psi(\vec{r},E)\rangle = \sum_n f_n(E)|f_n(\vec{r})\rangle$, where $\vec{r}$ is the configuration space coordinate. The basis functions

must be compatible with the domain of the Hamiltonian. They should also satisfy the boundary conditions. Typically (and especially when calculating the discrete spectrum) the choice of basis is limited to those that carry diagonal representations of the Hamiltonian. That is, one looks for an $L^2$ basis set $\left\{ \boldsymbol{f}_n \right\}_{n=0}^{\infty}$ such that $H \left| \boldsymbol{f}_n \right\rangle = E_n \left| \boldsymbol{f}_n \right\rangle$. The continuous spectrum is obtained from the analysis of an infinite sum of these *complete* basis functions. Truncating this sum, for numerical reasons, may create problems such as the presence of unphysical states or fictitious resonances in the spectrum.

In this article we relax the restriction of a diagonal matrix representation for the Hamiltonian. We only require that the hermitian matrix representation of the wave operator be tridiagonal. That is, the action of the wave operator on the elements of the basis is allowed to take the general form $(H-E)\left| \boldsymbol{f}_n \right\rangle \sim \left| \boldsymbol{f}_n \right\rangle + \left| \boldsymbol{f}_{n-1} \right\rangle + \left| \boldsymbol{f}_{n+1} \right\rangle$ such that

$$\left\langle \boldsymbol{f}_n \middle| H - E \middle| \boldsymbol{f}_m \right\rangle = (a_n - y)\boldsymbol{d}_{n,m} + b_n \boldsymbol{d}_{n,m-1} + b_{n-1}\boldsymbol{d}_{n,m+1} \tag{1.1}$$

where $y$ and the coefficients $\left\{ a_n, b_n \right\}_{n=0}^{\infty}$ are real and, in general, functions of the energy $E$, the angular momentum $\ell$, and potential parameters. Therefore, the matrix representation of the wave equation $(H-E)\left| \boldsymbol{y} \right\rangle = 0$, which is obtained by expanding $\left| \boldsymbol{y} \right\rangle$ as $\sum_m f_m \left| \boldsymbol{f}_m \right\rangle$ and projecting from the left on $\left\langle \boldsymbol{f}_n \right|$, results in the following three-term recursion relation

$$y f_n = a_n f_n + b_{n-1} f_{n-1} + b_n f_{n+1} \tag{1.2}$$

Consequently, the problem translates into finding solutions of the recursion relation for the expansion coefficients of the wavefunction $\boldsymbol{y}$. In most cases this recursion is solved easily and directly by correspondence with those for well known orthogonal polynomials. An example of a problem which is already solved using this approach is the Coulomb problem where the expansion coefficients of the wavefunction are written in terms of the Pollaczek polynomials [9]. Nonetheless, we find here another representation for the solution of the Coulomb problem which is written in terms of the continuous dual Hahn orthogonal polynomials [10]. In other problems, we found new recursion relations corresponding to new kind of orthogonal polynomials. Investigating the analytic solution of these recursion relations is a mathematical exercise which is more suitable for publication in a mathematically oriented journal elsewhere. Here, we find it adequate and physically sufficient to give a graphical representation of an approximation of the density (weight) function associated with some of these orthogonal polynomials for a given set of physical parameters. It should be noted that the solution of the problem as depicted by Eq. (1.2) above is obtained for all $E$, the discrete as well as the continuous. The representation equation (1.1) clearly shows that the discrete spectrum is easily obtained by diagonalization which requires that:

$$b_n = 0, \text{ and } a_n - y = 0 \tag{1.3}$$

Typically, there are two solutions of the recursion relation (1.2). One is regular and behaves asymptotically $(n \to \infty)$ as sine-like. The other is irregular and behaves asymptotically as cosine-like (i.e., it differs by a phase of $\boldsymbol{p}/2$ from the sine-like solutions). The scattering states and associated phase shifts could be obtained algebraically by studying these asymptotic limits.

In configuration space, with coordinate $x$, the wavefunction $\boldsymbol{y}_E(x)$ is expanded as $\sum_{n=0}^{\infty} f_n(E)\boldsymbol{f}_n(x)$ where the $L^2$ basis functions could be written as

$$\boldsymbol{f}_n(x) = A_n w_n(x) P_n(x) \tag{1.4}$$

$A_n$ is the normalization constant, $P_n(x)$ is a polynomial of degree $n$ in $x$, and the weight function satisfies $w_n(x_\pm) = 0$ where $x_-(x_+)$ is the left (right) boundary of configuration space. In the subsequent sections we consider examples in two spaces. One is where $x_\pm$ are finite and

$$w_n(x) = (x - x_-)^{\boldsymbol{a}}(x - x_+)^{\boldsymbol{b}}$$
$$P_n(x) = {}_2F_1(-n, b; c; x) \tag{1.5}$$

The other is semi-infinite where $x_-$ is finite, $x_+$ infinite, and where

$$w_n(x) = (x - x_-)^{\boldsymbol{a}} e^{-\boldsymbol{b} x}$$
$$P_n(x) = {}_1F_1(-n; c; x) \tag{1.6}$$

${}_2F_1$ is the hypergeometric function and ${}_1F_1$ is the confluent hypergeometric function. The parameters $\boldsymbol{a}, \boldsymbol{b}, b$ and $c$ are real with $\boldsymbol{a}$ and $\boldsymbol{b}$ positive. They are related to the physical parameters of the corresponding problem and may also depend on the index $n$.

In the following sections we consider various examples of problems in one and three dimensions. Solutions of some of the classic problems such as the Oscillator and Morse are reproduced adding, however, new tridiagonal representations. We also develop generalizations of others such as the Hulthén-type problems where we find new solutions and define their associated orthogonal polynomials. In addition, we investigate problems with hyperbolic potentials such as the Rosen-Morse type and present their generalized solutions. These investigations do not exhaust the set of all solvable problems using this approach. Furthermore, this development embodies powerful tools in the search for solutions of the wave equation by exploiting the intimate connection between tridiagonal matrices and the theory of orthogonal polynomials. In such analysis, one is at liberty to employ a wide range of well established methods and numerical techniques associated with these settings such as quadrature approximations and continued fractions [11]. On the other hand, little attention will given to mathematical rigor in the following development.

## II. THE COULOMB PROBLEM

We start by taking the configuration space coordinate as $x = \boldsymbol{l} r$, where $\boldsymbol{l}$ is a length scale parameter which is real and positive and $r$ is the radial coordinate in three dimensions. This problem belongs to the case described by Eq. (1.6) with $x_- = 0$. Since ${}_1F_1(-n; c; x)$ is proportional to the Laguerre polynomial $L_n^{c-1}(x)$ [12], we can write the basis functions as

$$\boldsymbol{f}_n(r) = A_n x^{\boldsymbol{a}} e^{-\boldsymbol{b} x} L_n^{\boldsymbol{n}}(x) \tag{2.1}$$

where $A_n = \sqrt{\boldsymbol{l}\, \Gamma(n+1)/\Gamma(n+\boldsymbol{n}+1)}$, $n = 0,1,2,\ldots$, $\boldsymbol{n} > -1$, $\boldsymbol{a}$ and $\boldsymbol{b}$ are real and positive. In atomic units ($\hbar = m = 1$), the radial time-independent Schrödinger wave equation for a structureless particle in a potential $V(r)$ reads

$$2(H - E)|\boldsymbol{y}\rangle = \left[ -\frac{d^2}{dr^2} + \frac{\ell(\ell+1)}{r^2} + 2V(r) - 2E \right]|\boldsymbol{y}\rangle = 0 \tag{2.2}$$

The action of the first term (the second order derivative) on the basis function results in the following

$$\frac{d^2 f_n}{dr^2} = \boldsymbol{l}^2 A_n x^a e^{-bx} \left[ \frac{d^2}{dx^2} + \left( \frac{2\boldsymbol{a}}{x} - 2\boldsymbol{b} \right) \frac{d}{dx} - \frac{\boldsymbol{a}}{x^2} + \left( \frac{\boldsymbol{a}}{x} - \boldsymbol{b} \right)^2 \right] L_n^{\boldsymbol{n}}(x) \qquad (2.3)$$

Using the differential equation and differential formula of the Laguerre polynomials [Eqs. (A.3) and (A.4) in the Appendix] we obtain the following action of the wave operator (2.2) on the basis

$$\frac{2}{\boldsymbol{l}^2}(H-E)\big|f_n\big\rangle = \left[ \frac{n}{x}\left( 2\boldsymbol{b} + \frac{\boldsymbol{n}+1-2\boldsymbol{a}}{x} \right) + \frac{\ell(\ell+1)-\boldsymbol{a}(\boldsymbol{a}-1)}{x^2} + \frac{2\boldsymbol{a}\boldsymbol{b}}{x} - \boldsymbol{b}^2 \right.$$
$$\left. + \frac{2}{\boldsymbol{l}^2}(V-E) \right]\big|f_n\big\rangle + \frac{n+\boldsymbol{n}}{x}\left( 1-2\boldsymbol{b} + \frac{2\boldsymbol{a}-\boldsymbol{n}-1}{x} \right)\frac{A_n}{A_{n-1}}\big|f_{n-1}\big\rangle \qquad (2.4)$$

The orthogonality relation for the Laguerre polynomials [shown in the Appendix as Eq. (A.5)] requires that $\boldsymbol{b} = \frac{1}{2}$ if we were to obtain a tridiagonal representation for $\big\langle f_n\big|H-E\big|f_m\big\rangle$. Moreover, we end up with only two possibilities to achieve the tridiagonal structure of the wave operator (2.4):

(1) $\boldsymbol{a} = \frac{\boldsymbol{n}+1}{2}$, $\boldsymbol{n} = \pm(2\ell+1)$, and $V = \frac{\boldsymbol{l}Z}{x} = \frac{Z}{r}$       (2.5)

(2) $\boldsymbol{a} = \frac{\boldsymbol{n}}{2}+1$, $\boldsymbol{l}^2 = -8E$, and $V = \frac{\boldsymbol{l}Z}{x} + \frac{\boldsymbol{l}^2 B}{x^2} = \frac{Z}{r} + \frac{B}{r^2}$       (2.6)

where $Z$ is the particle's charge number in units of $e$ and $B$ is a centripetal barrier potential parameter.

We start by considering the first possibility described by (2.5). Using the relation $L_n^{-m}(x) = (-x)^m \frac{(n-m)!}{n!} L_{n-m}^m(x)$ for $n \geq m \geq 1$, it is, therefore, sufficient to work out the case $\boldsymbol{n} = +(2\ell+1)$ only:

$$f_n(r) = A_n x^{\ell+1} e^{-x/2} L_n^{2(\ell+1)}(x) \qquad (2.7)$$

Substituting in Eq. (2.4) with $\boldsymbol{b} = \frac{1}{2}$ and projecting on $\big\langle f_m\big|$ we obtain

$$\frac{2}{\boldsymbol{l}^2}\big\langle f_n\big|H-E\big|f_m\big\rangle = \left[ 2(n+\ell+1)\left( \frac{1}{4} - \frac{2E}{\boldsymbol{l}^2} \right) + \frac{2Z}{\boldsymbol{l}} \right]d_{n,m}$$
$$+ \left( \frac{1}{4} + \frac{2E}{\boldsymbol{l}^2} \right)\left[ \sqrt{n(n+2\ell+1)}\,d_{n,m+1} + \sqrt{(n+1)(n+2\ell+2)}\,d_{n,m-1} \right] \qquad (2.8)$$

Therefore, the resulting recursion relation (1.2) for the expansion coefficients of the wavefunction becomes

$$\left[ 2\left( n+\ell+1+\frac{2Z}{\boldsymbol{l}} \right)\frac{\boldsymbol{s}_-}{\boldsymbol{s}_+} - \frac{4Z}{\boldsymbol{l}} \right]f_n - \sqrt{n(n+2\ell+1)}\,f_{n-1} - \sqrt{(n+1)(n+2\ell+2)}\,f_{n+1} = 0 \qquad (2.9)$$

where $\boldsymbol{s}_{\pm} = \frac{2E}{\boldsymbol{l}^2} \pm \frac{1}{4}$. Rewriting this recursion in terms of the polynomials $P_n(E) = \sqrt{\Gamma(n+2\ell+2)/\Gamma(n+1)}\,f_n(E)$, we obtain the more familiar recursion relation

$$\left[ 2\left( n+\ell+1+\frac{2Z}{\boldsymbol{l}} \right)\frac{\boldsymbol{s}_-}{\boldsymbol{s}_+} - \frac{4Z}{\boldsymbol{l}} \right]P_n - (n+2\ell+1)P_{n-1} - (n+1)P_{n+1} = 0 \qquad (2.10)$$

which is that of the Pollaczek polynomials [13] [see Eq. (A.11) in the Appendix]. Thus, we can write

$$f_n(E) = \sqrt{\frac{\Gamma(n+1)}{\Gamma(n+2\ell+2)}}\ \mathsf{P}_n^{\ell+1}\left( \frac{\boldsymbol{s}_-}{\boldsymbol{s}_+}; 1+\frac{2Z}{\boldsymbol{l}}, -\frac{2Z}{\boldsymbol{l}} \right) \qquad (2.11)$$

This solution of the Coulomb problem was obtained by Yamani and Reinhardt [9]. Restricting the representation (2.8) to the diagonal form gives the discrete spectrum via the requirement (1.3) which in this case reads as follows:

$$\frac{1}{4} + \frac{2E}{\boldsymbol{l}^2} = 0,$$
$$2(n+\ell+1)\left(\frac{1}{4} - \frac{2E}{\boldsymbol{l}^2}\right) + \frac{2Z}{\boldsymbol{l}} = 0 \tag{2.12}$$

This gives the following well known energy spectrum for the bound states of the Coulomb problem

$$E_n = -\frac{1}{2}\left(\frac{Z}{n+\ell+1}\right)^2, \quad \boldsymbol{l} = \boldsymbol{l}_n = 2|Z|/(n+\ell+1) \tag{2.13}$$

where $n = 0,1,2,\dots$.

For the second possibility described by (2.6) the basis functions are

$$\boldsymbol{f}_n(r) = A_n x^{1+\boldsymbol{n}/2} e^{-x/2} L_n^{\boldsymbol{n}}(x), \qquad x = 2\sqrt{-2E}\; r \tag{2.14}$$

Substituting in Eq. (2.4) and projecting on $\langle \boldsymbol{f}_m |$ we get

$$-\frac{1}{4E}\langle \boldsymbol{f}_n | H - E | \boldsymbol{f}_m \rangle = \left[(2n+\boldsymbol{n}+1)\left(n+\frac{\boldsymbol{n}}{2}+1+\boldsymbol{t}\right) - n - \left(\frac{\boldsymbol{n}+1}{2}\right)^2 + (\ell+\tfrac{1}{2})^2 + 2B\right]\boldsymbol{d}_{n,m}$$
$$-\left(n+\frac{\boldsymbol{n}}{2}+\boldsymbol{t}\right)\sqrt{n(n+\boldsymbol{n})}\,\boldsymbol{d}_{n,m+1} - \left(n+\frac{\boldsymbol{n}}{2}+1+\boldsymbol{t}\right)\sqrt{(n+1)(n+\boldsymbol{n}+1)}\,\boldsymbol{d}_{n,m-1} \tag{2.15}$$

where $\boldsymbol{t} = Z/\sqrt{-2E}$. The resulting recursion relation for the expansion coefficients of the wavefunction in terms of the polynomials $P_n(E) = \sqrt{\Gamma(n+1)/\Gamma(n+\boldsymbol{n}+1)}\, f_n(E)$ reads

$$\left[(2n+\boldsymbol{n}+1)\left(n+\frac{\boldsymbol{n}}{2}+1+\boldsymbol{t}\right) - n - \left(\frac{\boldsymbol{n}+1}{2}\right)^2 + (\ell+\tfrac{1}{2})^2 + 2B\right]P_n$$
$$- n\left(n+\frac{\boldsymbol{n}}{2}+\boldsymbol{t}\right)P_{n-1} - (n+\boldsymbol{n}+1)\left(n+\frac{\boldsymbol{n}}{2}+1+\boldsymbol{t}\right)P_{n+1} = 0 \tag{2.16}$$

This is a special case of the recursion relation for the continuous dual Hahn orthogonal polynomial [10] which is shown as Eq. (A.14) in the Appendix. As a result we obtain

$$f_n(E) = \sqrt{\frac{\Gamma(n+\boldsymbol{n}+1)}{\Gamma(n+1)}}\, S_n[-2B - (\ell+\tfrac{1}{2})^2; \frac{\boldsymbol{n}}{2}, \boldsymbol{t}+\tfrac{1}{2}, \frac{\boldsymbol{n}+1}{2}] \tag{2.17}$$

The discrete energy spectrum is evidently obtained by diagonalization of (2.15) which translates into the requirements

$$n + \frac{\boldsymbol{n}}{2} + 1 + \boldsymbol{t} = 0,$$
$$-\left(\frac{\boldsymbol{n}+1}{2}\right)^2 + (\ell+\tfrac{1}{2})^2 + 2B = 0 \tag{2.18}$$

giving,

$$E_n = -\frac{1}{2}\left(\frac{Z}{n+\boldsymbol{n}/2+1}\right)^2, \qquad \frac{\boldsymbol{n}+1}{2} = \pm\sqrt{(\ell+\tfrac{1}{2})^2 + 2B} \tag{2.19}$$

which is the same as that in Eq. (2.13) above when $B = 0$.

## III. THE SPHERICAL OSCILLATOR

In this case, we write the configuration space coordinate as $x = (\boldsymbol{l}\, r)^2$. Thus, the basis elements of the $L^2$ space is written as

$$\boldsymbol{f}_n(r) = A_n x^a e^{-b\,x} L_n^{\boldsymbol{n}}(x) \tag{3.1}$$

where again $\boldsymbol{a}$ and $\boldsymbol{b}$ are real and positive, $\boldsymbol{n} > -1$, and $A_n = \sqrt{2|\boldsymbol{l}|\,\Gamma(n+1)\big/\Gamma(n+\boldsymbol{n}+1)}$.
Using $\frac{d}{dr} = 2|\boldsymbol{l}|\sqrt{x}\,\frac{d}{dx}$, we obtain the following

$$\frac{d^2 \boldsymbol{f}_n}{dr^2} = 4\boldsymbol{l}^2 A_n x^{\boldsymbol{a}+1} e^{-bx} \left[ \frac{d^2}{dx^2} + \left( \frac{2\boldsymbol{a}+\frac{1}{2}}{x} - 2\boldsymbol{b} \right)\frac{d}{dx} - \frac{\boldsymbol{a}}{x^2} + \left( \frac{\boldsymbol{a}+\frac{1}{2}}{x} - \boldsymbol{b} \right)\left( \frac{\boldsymbol{a}}{x} - \boldsymbol{b} \right) \right] L_n^{\boldsymbol{n}}(x) \qquad (3.2)$$

The differential formulas for the Laguerre polynomials, which are shown in the Appendix, translates this equation into the following

$$\frac{d^2 \boldsymbol{f}_n}{dr^2} = 4\boldsymbol{l}^2 x \left[ \frac{n}{x}\left( \frac{2\boldsymbol{a}-\boldsymbol{n}-\frac{1}{2}}{x} - 2\boldsymbol{b} \right) + \frac{\boldsymbol{a}(\boldsymbol{a}-\frac{1}{2})}{x^2} - \frac{\boldsymbol{b}(2\boldsymbol{a}+\frac{1}{2})}{x} + \boldsymbol{b}^2 \right] \boldsymbol{f}_n$$

$$-4\boldsymbol{l}^2 (n+\boldsymbol{n})\left( \frac{2\boldsymbol{a}-\boldsymbol{n}-\frac{1}{2}}{x} + 1 - 2\boldsymbol{b} \right)\frac{A_n}{A_{n-1}} \boldsymbol{f}_{n-1} \qquad (3.3)$$

The tridiagonal structure can only be achieved if and only if $\boldsymbol{b} = \frac{1}{2}$ resulting in the following action of the wave operator on the basis

$$\frac{1}{2\boldsymbol{l}^2}(H-E)\big|\boldsymbol{f}_n\big\rangle = \left[ -\frac{n(2\boldsymbol{a}-\boldsymbol{n}-\frac{1}{2}) + (\boldsymbol{a}-\frac{1}{4})^2 - (\ell+\frac{1}{2})^2\big/4}{x} + n + \boldsymbol{a} + \frac{1}{4} - \frac{x}{4} \right.$$

$$\left. + \frac{1}{2\boldsymbol{l}^2}(V-E) \right]\big|\boldsymbol{f}_n\big\rangle + \frac{(n+\boldsymbol{n})(2\boldsymbol{a}-\boldsymbol{n}-\frac{1}{2})}{x}\frac{A_n}{A_{n-1}}\big|\boldsymbol{f}_{n-1}\big\rangle \qquad (3.4)$$

The orthogonality relation (A.5) and the tridiagonal requirement limit the possibilities to either one of the following two:

(1) $\boldsymbol{a} = \frac{\boldsymbol{n}+\frac{1}{2}}{2}$, $\boldsymbol{n} = \pm(\ell+\frac{1}{2})$, and $V = \frac{1}{2}\boldsymbol{w}^2 x = \frac{1}{2}(\boldsymbol{l}\,\boldsymbol{w})^2 r^2$ \qquad (3.5)

(2) $\boldsymbol{a} = \frac{\boldsymbol{n}+\frac{3}{2}}{2}$, and $V = \frac{1}{2}\boldsymbol{l}^2 x + \boldsymbol{l}^2 B\big/x = \frac{1}{2}\boldsymbol{l}^4 r^2 + B\big/r^2$ \qquad (3.6)

where $\boldsymbol{w}$ is the oscillator frequency and $B$ is a centripetal barrier potential parameter.

The first possibility where $\boldsymbol{n} = \pm(\ell+\frac{1}{2})$ gives the following tridiagonal matrix representation of the wave operator

$$\frac{2}{\boldsymbol{l}^2}\big\langle \boldsymbol{f}_n\big|H-E\big|\boldsymbol{f}_m\big\rangle = \left[ (2n+\boldsymbol{n}+1)\left( \frac{\boldsymbol{w}^2}{\boldsymbol{l}^2}+1 \right) - \frac{2E}{\boldsymbol{l}^2} \right]\boldsymbol{d}_{n,m}$$

$$-\left( \frac{\boldsymbol{w}^2}{\boldsymbol{l}^2}-1 \right)\left[ \sqrt{n(n+\boldsymbol{n})}\,\boldsymbol{d}_{n,m+1} + \sqrt{(n+1)(n+\boldsymbol{n}+1)}\,\boldsymbol{d}_{n,m-1} \right] \qquad (3.7)$$

Writing the resulting recursion relation in terms of the polynomials $P_n(E) = \sqrt{\Gamma(n+\boldsymbol{n}+1)\big/\Gamma(n+1)}\,f_n(E)$ gives the following

$$\left[ \left( 2n+\boldsymbol{n}+1-\frac{E}{\boldsymbol{l}^2} \right)\frac{\boldsymbol{s}_+}{\boldsymbol{s}_-} + \frac{E}{\boldsymbol{l}^2} \right]P_n - (n+\boldsymbol{n})\,P_{n-1} - (n+1)\,P_{n+1} = 0 \qquad (3.8)$$

where $\boldsymbol{s}_\pm = (\boldsymbol{w}/\boldsymbol{l})^2 \pm 1$. Comparing this with the recursion relation of the Pollaczek polynomial [Eq. (A.11) in the Appendix] we can directly write the expansion coefficients of the oscillator wavefunction as

$$f_n(E) = \sqrt{\frac{\Gamma(n+1)}{\Gamma(n+\boldsymbol{n}+1)}}\,\mathsf{P}_n^{\frac{\boldsymbol{n}+1}{2}}\left( \frac{\boldsymbol{s}_+}{\boldsymbol{s}_-}; 1 - \frac{E}{2\boldsymbol{l}^2}, \frac{E}{2\boldsymbol{l}^2} \right) \qquad (3.9)$$

The discrete energy spectrum is easily obtained from Eq. (3.7) by requiring that

$$\boldsymbol{l}^2 = \boldsymbol{w}^2, \text{ and } 2n+\boldsymbol{n}+1 = E\big/\boldsymbol{l}^2 \qquad (3.10)$$

giving,

$$E_n = \boldsymbol{w}^2\big(2n+\boldsymbol{n}+1\big) \qquad (3.11)$$

where $n = \pm(\ell + \frac{1}{2})$ and $n = 0,1,2,\ldots$.

Now, we consider the second possibility (3.6) in which the parameter $n$, aside from being $> -1$, is arbitrary. It is to be noted that in this case the first term in the potential, $\frac{1}{2}l^2 x$, is essential to cancel the contribution of the term $-\frac{x}{4}$ in Eq. (3.4) whose presence destroys the tridiagonal structure. Following the same procedure as outlined above we obtain the following matrix representation of the wave operator

$$\frac{1}{2l^2}\langle f_n | H - E | f_m \rangle = \left[ \left(2n+n+1\right)\left(n+\frac{n}{2}+1-t\right) - n - \left(\frac{n+1}{2}\right)^2 + \left(\frac{\ell+\frac{1}{2}}{2}\right)^2 + \frac{B}{2}\right] d_{n,m}$$
$$- \left(n+\frac{n}{2}-t\right)\sqrt{n(n+n)}\,d_{n,m+1} - \left(n+\frac{n}{2}+1-t\right)\sqrt{(n+1)(n+n+1)}\,d_{n,m-1} \qquad (3.12)$$

where $t = E/2l^2$. Similarly, we can easily show that the solution of the recursion relation (1.2) resulting from the matrix representation of the wave equation above could be written in terms of the continuous dual Hahn orthogonal polynomials as follows:

$$f_n(E) = \sqrt{\frac{\Gamma(n+n+1)}{\Gamma(n+1)}} S_n[-\frac{B}{2} - \left(\frac{\ell+\frac{1}{2}}{2}\right)^2 ; \frac{n+1}{2}, \frac{1}{2} - t, \frac{n+1}{2}] \qquad (3.13)$$

In addition, the discrete energy spectrum is obtained from (3.12) by the requirement that

$$n + \frac{n}{2} + 1 - t = 0,$$
$$- \left(\frac{n+1}{2}\right)^2 + \left(\frac{\ell+\frac{1}{2}}{2}\right)^2 + \frac{B}{2} = 0 \qquad (3.14)$$

Giving:

$$E_n = l^2 \left(2n+n+2\right), \quad n+1 = \pm\sqrt{\left(\ell+\frac{1}{2}\right)^2 + 2B} \qquad (3.15)$$

## IV. POWER-LAW POTENTIALS AT ZERO ENERGY

The configuration space coordinate for this problem is $x = (lr)^m$, where $l > 0$ and the real parameter $m \neq 0,1,2$. The three dismissed values of $m$ correspond to the Morse, Coulomb, and Oscillator problems, respectively. The elements of the basis wave functions are

$$f_n(r) = A_n x^a e^{-x/2} L_n^n(x) \qquad (4.1)$$

with $f_n(0) = f_n(\infty) = 0$ for positive or negative value of $m$, which is fixed once and for all. The normalization constant is $A_n = \sqrt{|m| l \Gamma(n+1)/\Gamma(n+n+1)}$. Using the derivative chain rule, which gives $\frac{d}{dr} = ml x^{1-1/m} \frac{d}{dx}$, we can write

$$\frac{2}{(ml)^2}\left(H - E\right)|f_n\rangle = x^{2-2/m}\left\{ \frac{1}{x^2}\left[\left(\frac{\ell+\frac{1}{2}}{m}\right)^2 - \left(a - \frac{1}{2m}\right)^2 - n\left(2a-n-\frac{1}{m}\right)\right]\right.$$
$$+ \frac{1}{x}\left(n+a+\frac{1}{2}-\frac{1}{2m}\right) - \frac{1}{4} + \frac{2}{(ml)^2} x^{-2+2/m}\left(V-E\right)\left.\right\}|f_n\rangle \qquad (4.2)$$
$$+ x^{-2/m}(n+n)\left(2a-n-\frac{1}{m}\right)\frac{A_n}{A_{n-1}}|f_{n-1}\rangle$$

The constant $E$ term must independently vanish if the representation of the wave operator is to be tridiagonal. Consequently, this case is analytically solvable only at zero energy. It is to be noted that this condition does not diminish the significance of these solutions. Zero energy solutions have valuable applications in scattering calculations (e.g., effective rang and scattering length parameters [14]) and in the investigation of low energy limit cases. For this problem we also end up with two possibilities for obtaining the tridiagonal structure of the wave operator:

(1) $a = \frac{n+1/m}{2}$, $n = \pm \frac{2\ell+1}{m}$, and $V = x^{-2/m}\left(Ax + \frac{1}{2}Bx^2\right)$  (4.3)

(2) $a = \frac{1+n+1/m}{2}$, and $V = x^{-2/m}\left[A + \frac{1}{2}Bx + \frac{1}{2}\left(ml/2\right)^2 x^2\right]$  (4.4)

where $A$ and $B$ are real potential parameters. The last term of the potential in the second case is necessary to eliminate the non-tridiagonal component coming from the contribution of the term $-\frac{1}{4}x^{2-2/m}$ in Eq. (4.2).

The resulting representation of the Hamiltonian for the first case (4.3) is

$$\frac{2}{(ml)^2}\langle f_n | H | f_m \rangle = \left\{\left[\frac{B}{(ml)^2} + \frac{1}{4}\right](2n+n+1) + \frac{2A}{(ml)^2}\right\}d_{n,m}$$
$$-\left[\frac{B}{(ml)^2} - \frac{1}{4}\right]\left[\sqrt{n(n+n)}d_{n,m+1} + \sqrt{(n+1)(n+n+1)}d_{n,m-1}\right]$$  (4.5)

The recursion relation resulting from this representation has the following solution in terms of the Pollaczek polynomials:

$$f_n(\ell) = \sqrt{\frac{\Gamma(n+1)}{\Gamma(n+n+1)}}\, P_n^{\frac{n+1}{2}}\left(\frac{s_+}{s_-}; 1 + \frac{2A}{(ml)^2}, -\frac{2A}{(ml)^2}\right)$$  (4.6)

where $s_\pm = B/(ml)^2 \pm 1/4$. On the other hand, the diagonal representation of the Hamiltonian is obtainable from Eq. (4.5) by the requirement that the potential parameters assume the following form:

$$B = \left(ml/2\right)^2, \text{ and } A = -\left(ml/2\right)^2\left(2n+1\mp\frac{2\ell+1}{m}\right)$$  (4.7)

This diagonal representation has already been obtained by this author [15] and others [16].

The second possibility defined in (4.4) produces the following tridiagonal representation for $H$

$$\frac{2}{(ml)^2}\langle f_n | H | f_m \rangle = \left[(2n+n+1)\left(n+\frac{n}{2}+1+t\right) - n - \left(\frac{n+1}{2}\right)^2 + \left(\frac{\ell+1/2}{m}\right)^2 + \frac{2A}{(ml)^2}\right]d_{n,m}$$
$$-\left(n+\frac{n}{2}+t\right)\sqrt{n(n+n)}d_{n,m+1} - \left(n+\frac{n}{2}+1+t\right)\sqrt{(n+1)(n+n+1)}d_{n,m-1}$$  (4.8)

where $n$ is positive but, otherwise, arbitrary and $t = B/(ml)^2$. The associated recursion relation is solved for the expansion coefficients of the wavefunction in terms of the continuous dual Hahn polynomials as follows:

$$f_n(E) = \sqrt{\frac{\Gamma(n+n+1)}{\Gamma(n+1)}}S_n\left[\frac{-2A}{(ml)^2} - \left(\frac{\ell+1/2}{m}\right)^2; \frac{n+1}{2}, t+\frac{1}{2}, \frac{n+1}{2}\right]$$  (4.9)

The diagonal representation is obtained by restricting the potential and basis parameters to satisfy

$$B = -(ml)^2\left(n+n/2+1\right),$$
$$\frac{n+1}{2} = \pm\sqrt{\left(\frac{\ell+1/2}{m}\right)^2 + \frac{2A}{(ml)^2}}$$  (4.10)

# V. HULTHÉN TYPE POTENTIALS

The Hulthén potential [17], which is written as $V(r) = -lZe^{-lr}/(1-e^{-lr})$, is used as a model for a screened Coulomb potential, where $l$ is the screening parameter. It is so, because for small $l$ we can write this potential as $V(r) \approx -\frac{Z}{r}e^{-lr}$. The configuration space coordinate which is compatible with these kind of problems is $x = 1 - 2e^{-lr}$. It maps real space into a bounded one. That is, $x \in [-1, +1]$ for $r \in [0, \infty]$. This problem belongs to the situation described by Eq. (1.5) with $x_\pm = \pm 1$. Since $_2F_1(-n, n+b; c; x)$ is proportional to the Jacobi polynomial $P_n^{(c-1, b-c)}(1-2x)$ [12], then the $L^2$ basis functions that satisfy the boundary conditions for this case could be written as

$$f_n(r) = A_n(1+x)^a(1-x)^b P_n^{(m,n)}(x) \tag{5.1}$$

where $a, b > 0$, $m, n > -1$ and the normalization constant is

$$A_n = \sqrt{\frac{l(2n+m+n+1)}{2^{m+n+1}} \frac{\Gamma(n+1)\Gamma(n+m+n+1)}{\Gamma(n+m+1)\Gamma(n+n+1)}} \tag{5.2}$$

Using the differential formulas of the Jacobi polynomials [Eqs. (A.8) and (A.9) in the Appendix], and $\frac{d}{dr} = l(1-x)\frac{d}{dx}$, we can write

$$\frac{d^2 f_n}{dr^2} = l^2 \frac{1-x}{1+x} \left\{ \left[ -n\left(x + \frac{n-m}{2n+m+n}\right)\left(\frac{m-2b}{1-x} + \frac{2a-n-1}{1+x}\right) - n(n+m+n+1) - a(2b+1) \right. \right.$$
$$\left. \left. + b^2 \frac{1+x}{1-x} + a(a-1)\frac{1-x}{1+x} \right] f_n + 2\frac{(n+m)(n+n)}{2n+m+n}\left(\frac{m-2b}{1-x} + \frac{2a-n-1}{1+x}\right)\frac{A_n}{A_{n-1}} f_{n-1} \right\} \tag{5.3}$$

Noting that $\int_0^\infty dr = \frac{1}{l}\int_{-1}^{+1}\frac{dx}{1-x}$ and using the orthogonality relation for the Jacobi polynomials [Eq. (A.10) in the Appendix], we arrive at the following conclusions. First, this problem admits only S-wave ($\ell = 0$) exact solutions since the orbital term creates intractable non-tridiagonal representations. Second, the tridiagonal requirement for the action of the wave operator limits the possibilities to the following three:

(1) $b = m/2$, and $a = (n+1)/2$ $\qquad\qquad$ (5.4)

(2) $b = m/2$, and $a = 1+n/2$ $\qquad\qquad$ (5.5)

(3) $b = (m+1)/2$, and $a = (n+1)/2$ $\qquad\qquad$ (5.6)

The first possibility eliminates the $f_{n-1}$ term from Eq. (5.3), whereas the last two allow this term to contribute to the matrix representation below and above the diagonal. The calculation in the first possibility (5.4) gives the following action of the S-wave Schrödinger operator on the basis:

$$\frac{2}{l^2}(H-E)|f_n\rangle = \frac{1-x}{1+x}\left[ n(n+m+n+1) + \frac{1}{2}(m+1)(n+1) \right.$$
$$\left. - \frac{m^2}{4}\frac{1+x}{1-x} - \frac{n^2-1}{4}\frac{1-x}{1+x} + \frac{2}{l^2}\frac{1+x}{1-x}(V-E) \right]|f_n\rangle \tag{5.7}$$

Therefore, to obtain the tridiagonal representation, our choice of potential functions is limited to those that satisfy the following constraint:

$$\frac{2}{l^2}\frac{1+x}{1-x}(V-E) = +\frac{m^2}{4}\frac{1+x}{1-x} + \frac{n^2-1}{4}\frac{1-x}{1+x} + \frac{2A}{l^2} + \frac{B}{l^2}(1\pm x) \tag{5.8}$$

where $A$ and $B$ are real potential parameters. The first two terms are necessary to cancel the contribution of the corresponding terms in Eq. (5.7) that destroy the tridiagonal structure. Equation (5.8) results in the following

$$E = -\frac{1}{2}\left(\lambda m/2\right)^2,\tag{5.9}$$

$$V = \frac{C}{(e^{\lambda r}-1)^2} + \frac{A}{e^{\lambda r}-1} + Be^{-\lambda r}\left\{\begin{array}{l}1\\ \frac{1}{e^{\lambda r}-1}\end{array}\right.\tag{5.10}$$

where $C = \frac{\lambda^2}{2}\frac{n^2-1}{4}$. The two alternatives in the last term of the potential correspond to the $\pm$ sign in Eq. (5.8). From now on, we will adopt the $+$ sign making the last potential term in (5.10) pure exponential, $Be^{-\lambda r}$. The first two terms in $V(r)$ are the Hulthén potential and its square. We obtain, after some manipulations, the following matrix representation of the wave operator

$$\frac{2}{\lambda^2}\left\langle f_n\left|H-E\right|f_m\right\rangle = \left[n\left(n+m+\nu+1\right) + \frac{2B}{\lambda^2}\frac{2n(n+m+\nu+1)+(m+\nu)(\nu+1)}{(2n+m+\nu)(2n+m+\nu+2)}\right.$$
$$\left.+\frac{1}{2}(m+1)(\nu+1) + \frac{2A}{\lambda^2}\right]d_{n,m} - \frac{2B/\lambda^2}{2n+m+\nu}\sqrt{\frac{n(n+m)(n+\nu)(n+m+\nu)}{(2n+m+\nu-1)(2n+m+\nu+1)}}d_{n,m+1}$$
$$+\frac{2B/\lambda^2}{2n+m+\nu+2}\sqrt{\frac{(n+1)(n+m+1)(n+\nu+1)(n+m+\nu+1)}{(2n+m+\nu+1)(2n+m+\nu+3)}}d_{n,m-1}\tag{5.11}$$

where $m = m(E)$ as given by Eq. (5.9). The resulting three-term recursion relation (1.2) could be written in terms of the polynomials defined by

$$P_n(E) = \frac{1}{\sqrt{2n+m+\nu+1}}\sqrt{\frac{\Gamma(n+m+1)\Gamma(n+\nu+1)}{\Gamma(n+1)\Gamma(n+m+\nu+1)}}f_n(E)\tag{5.12}$$

in which case it reads

$$yP_n = \left[\frac{g}{4}(2n+m+\nu+1)^2 + \frac{2n(n+m+\nu+1)+(m+\nu)(\nu+1)}{(2n+m+\nu)(2n+m+\nu+2)}\right]P_n$$
$$+\frac{(n+m)(n+\nu)}{(2n+m+\nu)(2n+m+\nu+1)}P_{n-1} + \frac{(n+1)(n+m+\nu+1)}{(2n+m+\nu+1)(2n+m+\nu+2)}P_{n+1}\tag{5.13}$$

where $g = \lambda^2/2B$ and $y(E) = (C-A-E)/B$. We are not aware of any known three-parameter orthogonal polynomials that satisfy the above recursion relation. However, comparing it to the recursion (A.6) in the Appendix suggests that these polynomials could be considered as a generalization or deformation of the Jacobi polynomials with $g$ being the deformation parameter ($g = 0$ corresponds to the Jacobi polynomial). Pursuing the analysis of these polynomials would be too mathematical and inappropriate for the present setting. It is being prepared for publication elsewhere. Nonetheless, we find it pressing to make the following remark: For large values of the index $n$ the $g$ term in the recursion (5.13) goes like $n^2$ whereas the rest of the terms go like $n^0$. Therefore, to obtain reasonable and meaningful numerical results, $g$ should be taken very small (i.e., $B \gg \lambda^2$). This means that the major contribution of the potential (5.10) in this case comes from the exponential component. Now back to the matrix representation (5.11) of the wave operator. The discrete energy spectrum is easily obtained by diagonalizing this representation which requires that $B = 0$ and

$$n\left(n+m+\nu+1\right) + \frac{1}{2}(m+1)(\nu+1) + 2A/\lambda^2 = 0\tag{5.14}$$

giving the following discrete spectrum for $n = 0,1,2,...$

$$E_n = -\frac{\lambda^2}{2}\left(\frac{m_n}{2}\right)^2 = -\frac{\lambda^2}{8}\left[n+\frac{\nu+1}{2} + \frac{2(A-C)/\lambda^2}{n+\frac{\nu+1}{2}}\right]^2\tag{5.15}$$

This agrees with the results obtained in Refs. [18].

Repeating the same analysis for the second possibility (5.5) and investigating the tridiagonal structure of the resulting action of the wave operator on the basis we conclude the following. First, the parameter $\boldsymbol{m}$ is related to the energy by Eq. (5.9) − the same way as in the first case. Second, the potential is required to take the following functional form

$$V = A\frac{1-x}{1+x} + 2B\frac{1-x}{(1+x)^2} = \frac{A}{e^{\boldsymbol{l}r}-1} + \frac{Be^{\boldsymbol{l}r}}{(e^{\boldsymbol{l}r}-1)^2} = \frac{A}{e^{\boldsymbol{l}r}-1} + \frac{B/4}{\sinh(\boldsymbol{l}r/2)^2} \tag{5.16}$$

The matrix representation of the S-wave Schrödinger operator is obtained as

$$\frac{1}{\boldsymbol{l}^2}\langle \boldsymbol{f}_n | H - E | \boldsymbol{f}_m \rangle =$$

$$\left\{ -y - \frac{n(n+\boldsymbol{m})}{2n+\boldsymbol{m}+\boldsymbol{n}} + \frac{2n(n+\boldsymbol{m}+\boldsymbol{n}+1)+(\boldsymbol{m}+\boldsymbol{n})(\boldsymbol{n}+1)}{(2n+\boldsymbol{m}+\boldsymbol{n})(2n+\boldsymbol{m}+\boldsymbol{n}+2)} \left[ \frac{1}{4}(2n+\boldsymbol{m}+\boldsymbol{n}+2)^2 + \boldsymbol{g} \right] \right\} d_{n,m}$$

$$+ \frac{1}{2n+\boldsymbol{m}+\boldsymbol{n}} \sqrt{\frac{n(n+\boldsymbol{m})(n+\boldsymbol{n})(n+\boldsymbol{m}+\boldsymbol{n})}{(2n+\boldsymbol{m}+\boldsymbol{n}-1)(2n+\boldsymbol{m}+\boldsymbol{n}+1)}} \left[ \frac{1}{4}(2n+\boldsymbol{m}+\boldsymbol{n})^2 + \boldsymbol{g} \right] d_{n,m+1} \tag{5.17}$$

$$+ \frac{1}{2n+\boldsymbol{m}+\boldsymbol{n}+2} \sqrt{\frac{(n+1)(n+\boldsymbol{m}+1)(n+\boldsymbol{n}+1)(n+\boldsymbol{m}+\boldsymbol{n}+1)}{(2n+\boldsymbol{m}+\boldsymbol{n}+1)(2n+\boldsymbol{m}+\boldsymbol{n}+3)}} \left[ \frac{1}{4}(2n+\boldsymbol{m}+\boldsymbol{n}+2)^2 + \boldsymbol{g} \right] d_{n,m-1}$$

where $y = \left(\frac{\boldsymbol{n}+1}{2}\right)^2 - 2B/\boldsymbol{l}^2 - 1/4$ and $\boldsymbol{g} = 2(E+A)/\boldsymbol{l}^2$. The resulting recursion relation in terms of the polynomials defined by Eq. (5.12) above reads as follows

$$yP_n = \left\{ -\frac{n(n+\boldsymbol{m})}{2n+\boldsymbol{m}+\boldsymbol{n}} + \frac{2n(n+\boldsymbol{m}+\boldsymbol{n}+1)+(\boldsymbol{m}+\boldsymbol{n})(\boldsymbol{n}+1)}{(2n+\boldsymbol{m}+\boldsymbol{n})(2n+\boldsymbol{m}+\boldsymbol{n}+2)} \left[ \frac{1}{4}(2n+\boldsymbol{m}+\boldsymbol{n}+2)^2 + \boldsymbol{g} \right] \right\} P_n$$

$$+ \frac{(n+\boldsymbol{m})(n+\boldsymbol{n})}{(2n+\boldsymbol{m}+\boldsymbol{n})(2n+\boldsymbol{m}+\boldsymbol{n}+1)} \left[ \frac{1}{4}(2n+\boldsymbol{m}+\boldsymbol{n})^2 + \boldsymbol{g} \right] P_{n-1} \tag{5.18}$$

$$+ \frac{(n+1)(n+\boldsymbol{m}+\boldsymbol{n}+1)}{(2n+\boldsymbol{m}+\boldsymbol{n}+1)(2n+\boldsymbol{m}+\boldsymbol{n}+2)} \left[ \frac{1}{4}(2n+\boldsymbol{m}+\boldsymbol{n}+2)^2 + \boldsymbol{g} \right] P_{n+1}$$

The three-parameter orthogonal polynomials defined by this recursion are, to the best of our knowledge, not seen before. The mathematical analysis of this recursion and the associated polynomials will not be carried out here. However, we find it appropriate and physically sufficient to plot the density (weight) function $\boldsymbol{r}_{\boldsymbol{g}}(y)$ which is used in the orthogonality relation of these polynomials. We use one of three numerical methods developed in Ref. [19] to obtain highly accurate approximations of the density function associated with a finite tridiagonal matrix. Figure (1) shows this density function for a given value of $\boldsymbol{m}$ and $\boldsymbol{n}$ and for several choices of the parameter $\boldsymbol{g}$. The discrete energy spectrum is obtained (by diagonalization) from Eq. (5.17) as

$$E_n = -\frac{\boldsymbol{l}^2}{8} \left[ n + \frac{\boldsymbol{n}+1}{2} + 1 + \frac{2A/\boldsymbol{l}^2}{n + \frac{\boldsymbol{n}+1}{2} + 1} \right]^2, \qquad n = 0,1,2,... \tag{5.19}$$

where $\frac{\boldsymbol{n}+1}{2} = \pm\left(\frac{C}{\boldsymbol{l}} + \frac{1}{2}\right)$ and $C$ is defined by $2B = C(C+\boldsymbol{l})$.

We leave it to the reader to find the recursion relation for the third possibility (5.6) and to verify the following results:

$$V = \frac{C}{(e^{\boldsymbol{l}r}-1)^2} + \frac{A}{e^{\boldsymbol{l}r}-1} + \frac{Be^{\boldsymbol{l}r}}{e^{\boldsymbol{l}r}-1} \tag{5.20}$$

where $C = \frac{\boldsymbol{l}^2}{2}\frac{\boldsymbol{n}^2-1}{4}$, $A$ and $B$ are real potential parameters. Moreover,

$$\frac{1}{l^2}\langle f_n | H - E | f_m \rangle =$$

$$\left\{ -y - \frac{n(n+\pmb{n})}{2n+\pmb{m}+\pmb{n}} + \frac{2n(n+\pmb{m}+\pmb{n}+1) + (\pmb{m}+\pmb{n})(\pmb{m}+1)}{(2n+\pmb{m}+\pmb{n})(2n+\pmb{m}+\pmb{n}+2)} \left[ \frac{1}{4}(2n+\pmb{m}+\pmb{n}+2)^2 + \pmb{g} \right] \right\} \pmb{d}_{n,m}$$

$$-\frac{1}{2n+\pmb{m}+\pmb{n}} \sqrt{\frac{n(n+\pmb{m})(n+\pmb{n})(n+\pmb{m}+\pmb{n})}{(2n+\pmb{m}+\pmb{n}-1)(2n+\pmb{m}+\pmb{n}+1)}} \left[ \frac{1}{4}(2n+\pmb{m}+\pmb{n})^2 + \pmb{g} \right] \pmb{d}_{n,m+1}$$ 

$$-\frac{1}{2n+\pmb{m}+\pmb{n}+2} \sqrt{\frac{(n+1)(n+\pmb{m}+1)(n+\pmb{n}+1)(n+\pmb{m}+\pmb{n}+1)}{(2n+\pmb{m}+\pmb{n}+1)(2n+\pmb{m}+\pmb{n}+3)}} \left[ \frac{1}{4}(2n+\pmb{m}+\pmb{n}+2)^2 + \pmb{g} \right] \pmb{d}_{n,m-1}$$

(5.21)

Aside from some sign changes and the exchange $\pmb{m} \leftrightarrow \pmb{n}$, this is the same as Eq. (5.17) above. However, here we have $y = \frac{2}{l^2}(E - B) + \left( \frac{\pmb{m}+1}{2} \right)^2$ and $\pmb{g} = 2(E + A - C)/l^2$. The discrete energy spectrum is obtained as

$$E_n = B - \frac{l^2}{2}\left( \frac{\pmb{m}+1}{2} \right)^2 = B - \frac{l^2}{8}\left[ n + \frac{\pmb{n}+1}{2} + \frac{2(A+B-C)/l^2}{n + \frac{\pmb{n}+1}{2}} \right]^2$$

(5.22)

In the examples of the following two sections we only list the results without giving details of the calculation.

## VI. THE S-WAVE MORSE POTENTIAL

$$x = \pmb{m}e^{-lr} ; \quad \pmb{m}, l > 0 \tag{6.1}$$

$$f_n(r) = A_n x^{\pmb{a}} e^{-x/2} L_n^{\pmb{n}}(x) , \quad A_n = \sqrt{l\,\Gamma(n+1)/\Gamma(n+\pmb{n}+1)} \tag{6.2}$$

Case (1): $\pmb{a} = \pmb{n}/2$

$$V = \pmb{m} A e^{-lr} + \pmb{m}^2 B e^{-2lr} \tag{6.3}$$

$$\frac{2}{l^2}\langle f_n | H - E | f_m \rangle = \left[ (2n+\pmb{n}+1)\left( \frac{2B}{l^2} + \frac{1}{4} \right) + \frac{2A}{l^2} \right] \pmb{d}_{n,m}$$

$$-\left( \frac{2B}{l^2} - \frac{1}{4} \right) \left[ \sqrt{n(n+\pmb{n})}\,\pmb{d}_{n,m+1} + \sqrt{(n+1)(n+\pmb{n}+1)}\,\pmb{d}_{n,m-1} \right]$$

(6.4)

where $\pmb{n} = \frac{2}{l}\sqrt{-2E}$ .

$$f_n(E) = \sqrt{\frac{\Gamma(n+1)}{\Gamma(n+\pmb{n}+1)}} \; P_n^{\frac{\pmb{n}+1}{2}}\left( \frac{\pmb{s}_+}{\pmb{s}_-}; 1 + \frac{2A}{l^2}, -\frac{2A}{l^2} \right) \tag{6.5}$$

where $\pmb{s}_\pm = 2B/l^2 \pm 1/4$. The discrete energy spectrum requirement puts $B = \frac{1}{2}\left( l/2 \right)^2$ and gives

$$E_n = -\frac{l^2}{2}\left( \frac{2A}{l^2} + n + \frac{1}{2} \right)^2 \tag{6.6}$$

Case (2): $\pmb{a} = (\pmb{n}+1)/2$

$$V = \pmb{m} A e^{-lr} + \frac{1}{2}\left( \pmb{m}l/2 \right)^2 e^{-2lr} \tag{6.7}$$

$$\frac{2}{l^2}\langle f_n | H - E | f_m \rangle = \left[ (2n+\pmb{n}+1)\left( n + \frac{\pmb{n}}{2} + 1 + \frac{2A}{l^2} \right) - n - \left( \frac{\pmb{n}+1}{2} \right)^2 - \frac{2E}{l^2} \right] \pmb{d}_{n,m}$$

$$-\left( n + \frac{\pmb{n}}{2} + \frac{2A}{l^2} \right) \sqrt{n(n+\pmb{n})}\,\pmb{d}_{n,m+1} - \left( n + \frac{\pmb{n}}{2} + 1 + \frac{2A}{l^2} \right) \sqrt{(n+1)(n+\pmb{n}+1)}\,\pmb{d}_{n,m-1}$$

(6.8)

$$f_n(E) = \sqrt{\frac{\Gamma(n+\pmb{n}+1)}{\Gamma(n+1)}} S_n[\frac{2E}{l^2}; \frac{\pmb{n}+1}{2}, \frac{2A}{l^2} + \frac{1}{2}, \frac{\pmb{n}+1}{2}] \tag{6.9}$$

This solution coincides with the findings in Refs. [20].

$$E_n = -\frac{l^2}{2}\left(\frac{n+1}{2}\right)^2 = -\frac{l^2}{2}\left(\frac{2A}{l^2}+n+\frac{1}{2}\right)^2 \tag{6.10}$$

## VII. ROSEN-MORSE TYPE POTENTIALS

$$f_n(x) = A_n(1+z)^a(1-z)^b P_n^{(m,n)}(z) \tag{7.1}$$

where $z = \tanh(l\,x)$ and $x \in \Re^1$. $a, b, m, n, l$ are real parameters with $a$, $b$ and $l$ positive. The normalization constant is as given by Eq. (5.2). Analytic solutions of this problem are obtainable for three cases where the parameters are related as: $(a, b) = \left(\frac{n}{2}, \frac{m}{2}\right)$, $\left(\frac{n}{2}, \frac{m+1}{2}\right)$, or $\left(\frac{n+1}{2}, \frac{m}{2}\right)$. As an example, we consider only the first case where the potential function assumes the following form which is compatible with the tridiagonal representation

$$V = C\tanh(l\,x) + \frac{A}{\cosh(l\,x)^2} + \frac{B/2}{\cosh(l\,x)^2}\left[1 \pm \tanh(l\,x)\right] \tag{7.2}$$

where $C = (l\,m/2)^2 - (l\,n/2)^2$. We consider, in what follows, the potential with the top $+$ sign in (7.2). The tridiagonal representation of the wave operator becomes

$$\begin{aligned}
\frac{2}{l^2}\langle f_n | H - E | f_m\rangle &= \left[\frac{1}{4}(2n+m+n+1)^2 + \frac{2B}{l^2}\frac{2n(n+m+n+1)+(m+n)(n+1)}{(2n+m+n)(2n+m+n+2)} + \frac{2A}{l^2} - \frac{1}{4}\right]d_{n,m} \\
&+ \frac{2B/l^2}{2n+m+n}\sqrt{\frac{n(n+m)(n+n)(n+m+n)}{(2n+m+n-1)(2n+m+n+1)}}d_{n,m+1} \\
&+ \frac{2B/l^2}{2n+m+n+2}\sqrt{\frac{(n+1)(n+m+1)(n+n+1)(n+m+n+1)}{(2n+m+n+1)(2n+m+n+3)}}d_{n,m-1}
\end{aligned} \tag{7.3}$$

The resulting recursion relation is similar to (5.13). The discrete energy spectrum is obtainable only for $B = 0$ which corresponds to the hyperbolic Rosen-Morse potential [21]. In this case we obtain:

$$E_n = -(l\,m/2)^2 - (l\,n/2)^2 = -\frac{l^2}{2}\left[(D/l - n)^2 + \left(\frac{C}{l^2}\right)^2 (D/l - n)^{-2}\right] \tag{7.4}$$

where $D$ is defined by $D(D+l) = -2A$.

It should finally be noted that the examples presented in this work do not exhaust all possible potentials in this larger class of analytically solvable systems. Moreover, this approach could be easily extended to the study of quasi exactly and conditionally exactly solvable problems. In addition, we believe that this development could also be extended to relativistic problems as well.

## ACKNOWLEDGEMENT

The author is grateful to M.E.H. Ismail for the help in identifying the orthogonal polynomials associated with some of the recursion relations in this work.

APPENDIX

The following are useful formulas and relations satisfied by the orthogonal polynomials that are relevant to the developments carried out in this work. They are found on most textbooks on orthogonal polynomials [12]. We list them here for ease of reference.

(1) The Laguerre polynomials $L_n^n(x)$:

$$xL_n^n = (2n + \boldsymbol{n} + 1)L_n^n - (n + \boldsymbol{n})L_{n-1}^n - (n+1)L_{n+1}^n \tag{A.1}$$

$$L_n^n(x) = \frac{\Gamma(n+1)\,\Gamma(\boldsymbol{n}+1)}{\Gamma(n+\boldsymbol{n}+1)}\,{}_1F_1(-n; \boldsymbol{n}+1;\, x) \tag{A.2}$$

$$\left[x\frac{d^2}{dx^2} + \left(\boldsymbol{n}+1-x\right)\frac{d}{dx} + n\right]L_n^n(x) = 0 \tag{A.3}$$

$$x\frac{d}{dx}L_n^n = nL_n^n - (n+\boldsymbol{n})L_{n-1}^n \tag{A.4}$$

$$\int_0^\infty x^n e^{-x} L_n^n(x) L_m^n(x)\,dx = \frac{\Gamma(n+\boldsymbol{n}+1)}{\Gamma(n+1)}\,\boldsymbol{d}_{nm} \tag{A.5}$$

(2) The Jacobi polynomials $P_n^{(\boldsymbol{m},\boldsymbol{n})}(x)$:

$$\left(\frac{1\pm x}{2}\right)P_n^{(\boldsymbol{m},\boldsymbol{n})} = \frac{2n(n+\boldsymbol{m}+\boldsymbol{n}+1)+(\boldsymbol{m}+\boldsymbol{n})(\frac{\boldsymbol{m}+\boldsymbol{n}}{2}\pm\frac{\boldsymbol{n}-\boldsymbol{m}}{2}+1)}{(2n+\boldsymbol{m}+\boldsymbol{n})(2n+\boldsymbol{m}+\boldsymbol{n}+2)}P_n^{(\boldsymbol{m},\boldsymbol{n})}$$
$$\pm\frac{(n+\boldsymbol{m})(n+\boldsymbol{n})}{(2n+\boldsymbol{m}+\boldsymbol{n})(2n+\boldsymbol{m}+\boldsymbol{n}+1)}P_{n-1}^{(\boldsymbol{m},\boldsymbol{n})} \pm \frac{(n+1)(n+\boldsymbol{m}+\boldsymbol{n}+1)}{(2n+\boldsymbol{m}+\boldsymbol{n}+1)(2n+\boldsymbol{m}+\boldsymbol{n}+2)}P_{n+1}^{(\boldsymbol{m},\boldsymbol{n})} \tag{A.6}$$

$$P_n^{(\boldsymbol{m},\boldsymbol{n})}(x) = \frac{\Gamma(n+\boldsymbol{m}+1)}{\Gamma(n+1)\,\Gamma(\boldsymbol{m}+1)}\,{}_2F_1(-n, n+\boldsymbol{m}+\boldsymbol{n}+1;\,\boldsymbol{m}+1;\,\tfrac{1-x}{2}) \tag{A.7}$$

$$\left\{\left(1-x^2\right)\frac{d^2}{dx^2} - \left[\left(\boldsymbol{m}+\boldsymbol{n}+2\right)x+\boldsymbol{m}-\boldsymbol{n}\right]\frac{d}{dx} + n\left(n+\boldsymbol{m}+\boldsymbol{n}+1\right)\right\}P_n^{(\boldsymbol{m},\boldsymbol{n})}(x) = 0 \tag{A.8}$$

$$\left(1-x^2\right)\frac{d}{dx}P_n^{(\boldsymbol{m},\boldsymbol{n})} = -n\left(x+\frac{\boldsymbol{n}-\boldsymbol{m}}{2n+\boldsymbol{m}+\boldsymbol{n}}\right)P_n^{(\boldsymbol{m},\boldsymbol{n})} + 2\frac{(n+\boldsymbol{m})(n+\boldsymbol{n})}{2n+\boldsymbol{m}+\boldsymbol{n}}P_{n-1}^{(\boldsymbol{m},\boldsymbol{n})} \tag{A.9}$$

$$\int_{-1}^{+1}(1-x)^{\boldsymbol{m}}(1+x)^{\boldsymbol{n}}P_n^{(\boldsymbol{m},\boldsymbol{n})}(x)P_m^{(\boldsymbol{m},\boldsymbol{n})}(x)\,dx = \frac{2^{\boldsymbol{m}+\boldsymbol{n}+1}}{2n+\boldsymbol{m}+\boldsymbol{n}+1}\frac{\Gamma(n+\boldsymbol{m}+1)\Gamma(n+\boldsymbol{n}+1)}{\Gamma(n+1)\Gamma(n+\boldsymbol{m}+\boldsymbol{n}+1)}\,\boldsymbol{d}_{nm} \tag{A.10}$$

(3) The Pollaczek polynomials $\mathsf{P}_n^n(x;a,b)$:

$$2\left[(n+\boldsymbol{n}+a-1)x+b\right]\mathsf{P}_n^n - (n+2\boldsymbol{n}-1)\mathsf{P}_{n-1}^n - (n+1)\mathsf{P}_{n+1}^n = 0 \tag{A.11}$$

$$\mathsf{P}_n^n(x;a,b) = \frac{\Gamma(n+2\boldsymbol{n})}{\Gamma(n+1)\Gamma(2\boldsymbol{n})}\,e^{in\boldsymbol{q}}\,{}_2F_1(-n, \boldsymbol{n}+i\boldsymbol{w}; 2\boldsymbol{n}; 1-e^{-2i\boldsymbol{q}}) \tag{A.12}$$

where $\boldsymbol{w} = (ax+b)\big/\sqrt{1-x^2}$ and $x = \cos\boldsymbol{q}$

$$\int_{-1}^{+1}\boldsymbol{r}^n(x)\mathsf{P}_n^n(x)\mathsf{P}_m^n(x)\,dx = \frac{\Gamma(n+2\boldsymbol{n})}{\Gamma(n+1)}\frac{\boldsymbol{p}/2^{2\boldsymbol{n}-1}}{n+\boldsymbol{n}+a}\,\boldsymbol{d}_{nm} \tag{A.13}$$

where $\boldsymbol{r}^n(x) = (\sin\boldsymbol{q})^{2\boldsymbol{n}-1}e^{(2\boldsymbol{q}-\boldsymbol{p})\boldsymbol{w}}\left|\Gamma(\boldsymbol{n}+i\boldsymbol{w})\right|^2$

(4) The continuous dual Hahn polynomials $\mathsf{S}_n(x^2;a,b,c)$:

$$x^2 \mathsf{S}_n = \Big[ (n+a+b)(n+a+c) + n(n+b+c-1) - a^2 \Big] \mathsf{S}_n$$
$$-n(n+b+c-1)\mathsf{S}_{n-1} - (n+a+b)(n+a+c)\mathsf{S}_{n+1} \tag{A.14}$$

$$\mathsf{S}_n(x^2; a, b, c) = {}_3F_2\left( \begin{matrix} -n, a+ix, a-ix \\ a+b, a+c \end{matrix} \middle| 1 \right) \tag{A.15}$$

$$\int_0^\infty \boldsymbol{r}(x)\mathsf{S}_n(x^2)\mathsf{S}_m(x^2)\,dx = \frac{\Gamma(n+1)\Gamma(n+b+c)}{\Gamma(n+a+b)\Gamma(n+a+c)}\boldsymbol{d}_{nm} \tag{A.16}$$

where $\boldsymbol{r}(x) = \dfrac{1}{2\boldsymbol{p}}\left| \dfrac{\Gamma(a+ix)\Gamma(b+ix)\Gamma(c+ix)}{\Gamma(a+b)\Gamma(a+c)\Gamma(2ix)} \right|^2 .$

Fig. (1): A graph of the density (weight) function $r_g(y)$ associated with the three-parameter orthogonal polynomials satisfying the recursion relation (5.18). The "Dispersion Correction" method developed in Ref. [19] is used in generating this plot using the recursion coefficients $\{a_n, b_n\}_{n=0}^{50}$ in Eq. (5.17) with $m = 1.0$, $n = 1.5$ and for several values of the parameter $g$ shown on the traces.

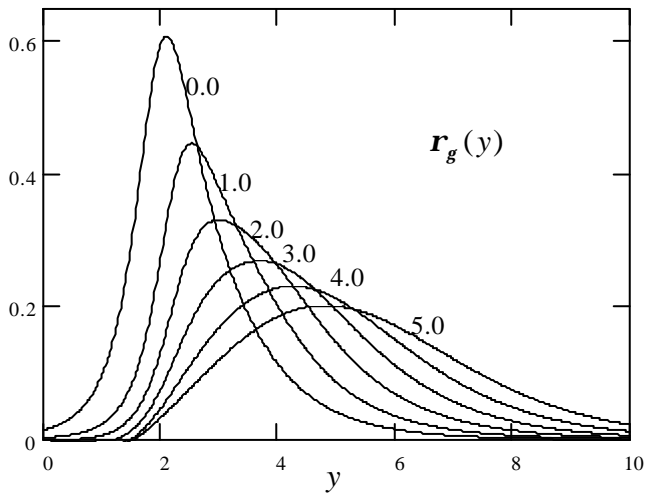

Fig. 1